\documentclass[12pt]{article}
\usepackage{amsmath}
\usepackage{amssymb}
\usepackage[dvips]{graphicx}

\numberwithin{equation}{section}

\newcommand{\CF}{ {\cal F} }
\newcommand{\CG}{ {\cal G} }
\newcommand{\CN}{ {\cal N} }
\newcommand{\CZ}{ {\cal Z} }

\newcommand{\hv}{ h^{\vee}  }
\newcommand{\av}{ \alpha^{\vee}  }
\newcommand{\AD}{ A_{D}  }

\begin{document}

\addtolength{\baselineskip}{2pt}
\thispagestyle{empty}

\vspace{2.5cm}


\begin{center}
{\scshape\large 
simple look at
\\
supersymmetric $\CN=2$ gauge theory
\\

with arbitrary gauge group}

\vspace{1cm}

{\scshape\large  Michael Yu. Kuchiev}

\vspace{1.5cm}
{\sl School of Physics, University of New South Wales,\\
Sydney, Australia}\\
{\tt kmy@phys.unsw.edu.au}\\\vspace{1cm}

{\Large ABSTRACT}
\vspace{0.3cm}

\end{center}

A new discrete symmetry is shown to govern and simplify low-energy properties of the supersymmetric $\CN=2$ gauge theory with an arbitrary  gauge group.  Each element of the related symmetry group $S_r$, $r$ being the rank of the gauge group, represents a permutation of $r$ electric charges available in the theory accompanied by a concurrent permutation of $r$ monopoles, provided the sets of charges and monopoles are chosen properly. The superpotential is symmetric under $S_r$. This symmetry strongly manifests itself for the degenerate case; when the masses of $r$ electric charges are chosen to be equal, then the masses of $r$ monopoles are necessarily degenerate as well, and vice versa. This condition uniquely defines the vital for the theory  VEV of the scalar field, which makes all monopoles massless. 
\newpage

\date{today}

\section{Introduction}
\label{intro}

The Seiberg-Witten problem for the supersymmetric $\CN=2$ gauge theory with an arbitrary gauge group is shown to possess an additional discrete symmetry compared to the simplest case when the gauge group is SU(2). 

Seiberg and  Witten \cite{Seiberg:1994rs,Seiberg:1994aj} explained that the low-energy properties of the supersymmetric $\CN=2$ gauge theory with SU(2) gauge group
can be formulated in a closed, explicit form.  Their discovery inspired  discussion of a more general case, call it the general Seiberg-Witten problem, in which low-energy properties of the supersymmetric $\CN=2$ theory with an arbitrary gauge group are revealed in a closed explicit form.

The search for the solution of this Seiberg-Witten problem was originated by
Klemm {\em et al} \cite{Klemm:1994qs} and Argyres and Faraggi
\cite{Argyres:1994xh}, who proposed the model for SU($n$) gauge groups. The models with other classical gauge groups were later presented in a number of works; brief reviews and references are given in \cite{D'Hoker:1999ft,Wu2006503}.
More references can be found in \cite{Kuchiev:2008mv}. 
However, in the latter work there had been identified the difficulty in the theory. Different models and techniques were employed even for theories with the classical gauge groups; for exceptional groups the situation looked even more complicated. As a result it was not clear whether the diversity of approaches reflects the nature of the Seiberg-Witten problem,  or it is artificially introduced by different models. 

Aiming to clarity this point Ref. \cite{Kuchiev:2008mv} found
an additional restriction on the theory stating that there exists 
the set of ``light" dyons; each of them becomes massless when the state of the theory is specifically chosen. 
Technical problems made it difficult to verify whether the available models complied with this condition or not, but
it was inspirational for Ref. \cite{Kuchiev:2009ez}, which formulated the general model valid for the supersymmetric $\CN=2$ gauge theory with an arbitrary compact simple gauge group, classical or exceptional. Technical problems (again) made direct comparison of this solution with the previously proposed ones cumbersome.

The situation was elucidated in \cite{Kuchiev:2011vh}, which argued that for gauge groups of rank $r>1$ the theory possesses the discrete symmetry under permutations within the specific set of dyons. For the SU(2) gauge group, which has $r=1$, the permutation group in question is trivial and hence immaterial to the Seiberg-Witten solution \cite{Seiberg:1994rs,Seiberg:1994aj}.
The solution of the general Seiberg-Witten problem proposed in \cite{Kuchiev:2009ez} complies with the symmetry condition formulated in \cite{Kuchiev:2011vh}, which distinguishes it from the previously considered approaches where this symmetry is not reproduced.

\section{Discrete symmetries}
\label{Dyons}
Consider the supersymmetric $\CN=2$ gauge theory. It was pointed out in \cite{Seiberg:1994rs,Seiberg:1994aj} that its low-energy properties are restricted by several conditions including the positive sign of the imaginary part of the effective coupling constant \cite{Witten:1979ey}, the holomorphicity of the superpotential \cite{Seiberg:1988ur},
symmetry under discrete chiral transformations, 
the value of the Witten index \cite{Witten:1982df,Witten:1997bs}, 
and duality. 

Putting together appropriately these restrictions allow one to define the explicit low-energy solution for the theory with the SU(2)  gauge group. Generalizing the problem for gauge groups with $r>1$ one needs to consider whether the same restrictions remain sufficient for identifying the solution in this case.  It turns out that  
for $r>1$ there exists an additional condition related to the specific discrete symmetry, which is formulated below, see Eqs. 
(\ref{Fsym})-(\ref{permD}) and (\ref{FDsym}), 
after relevant known facts are briefly outlined.

\subsection{Basic properties}
The theory is governed by the vacuum expectation value (VEV) of the scalar field, which belongs to the Cartan subalgebra of the gauge algebra and can be presented as an $r$-dimensional vector $A$. Similarly the VEV of the dual scalar field is also an $r$-dimensional vector $A_D$. This makes the electric $q$ and magnetic $g$ charges of massive dyons available in the theory $r$-dimensional vectors as well. 

\begin{figure}[tbh]
  \centering \includegraphics[ height=5.5cm, keepaspectratio=true, angle=0]{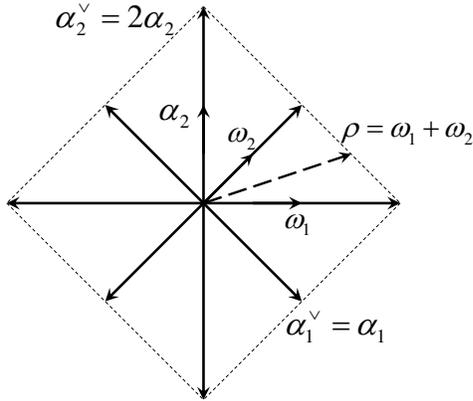}
\vspace{0cm}
\caption{ The Cartan algebra of SO(5) group: the simple roots $\alpha_1$ and $\alpha_2$, simple coroots $\av_1$  and $\av_2$,
and fundamental weights $\omega_1$ and $\omega_2$, as well as the Cartan vector $\rho$, see (\ref{ro}), are shown; thin dotted lines are drawn to clarify relative sizes of different vectors.}
\label{SO5} 
   \end{figure}

\begin{figure}[tbh]
  \centering \includegraphics[ height=5cm, keepaspectratio=true, angle=0]{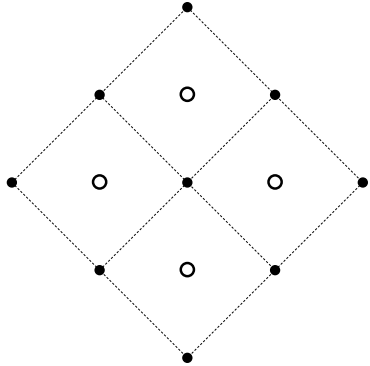}
\vspace{0cm}
\caption{Black dots - vertexes of the lattice of coroots $\mathbb  Q^\vee$ of the Cartan algebra for SO(5) group; empty circles - additional vertexes, which together with the black ones are present in the lattice of roots $\mathbb Q\supset \mathbb Q^\vee$.}
 \label{SO5-Q} 
   \end{figure}

It is known, see e. g.  \cite{Weinberg:2006rq}, that the Dirac-Schwinger-Zwanziger quantization conditions in general case ensure that the electric charges belong to the lattice of roots ${\mathbb Q}$,  while magnetic ones to the lattice of coroots ${\mathbb Q}^\vee$ of the Cartan algebra.
(For properties of the Lie algebras see e. g. the book \cite{Di-Francesco:1997}. The basic general ideas are illustrated in Figs. \ref{SO5} and
\ref{SO5-Q}, where properties of the Cartan algebra of the SO(5) group are depicted.)
   \noindent

In the supersymmetric $\CN=2$ gauge theory there is a simplification. The discrete chiral transformations and duality ensure that electric and magnetic charges belong to the same lattice. Consequently both sets of charges have to occupy vertexes of the same lattice ${\mathbb Q}^\vee\subset \mathbb Q$
\begin{equation}
q\,=\,\sum_{i=1}^r\,n^{(q)}_i\,\av_i~,
\quad\quad
g\,=\,\sum_{i=1}^r\,n^{(g)}_i\,\av_i~.
\label{g=ng}
\end{equation}
Here $\av_i$ are $r$ simple coroots of the 
Cartan algebra from which the lattice ${\mathbb Q}^\vee$ is constructed, see $\av_1,\,\av_2$ in Fig. \ref{SO5} and $\mathbb Q^\vee$ in Fig. \ref{SO5-Q},
while $n^{(q)}_i$ and $n^{(g)}_i$ are integers, 
$n^{(q)}_i,\,n^{(g)}_i\in \mathbb Z$. 
Fig. \ref{SO5-Q} illustrates the fact that generically ${\mathbb Q}^\vee\subset \mathbb Q$, though for simply laced groups ${\mathbb Q}^\vee = \mathbb Q$.
The scalar products $\av_i\cdot \av_j\in \mathbb{Z}$ are integer-valued. As a result scalar products for any electric and magnetic charges  satisfying (\ref{g=ng}) are also integer-valued, $q\cdot g\in \mathbb Z$, in accord with the quantization condition.  
   \noindent

Eqs. (\ref{g=ng})  show that we can conveniently describe 
electric and magnetic charges in the basis of simple coroots  $\av_i$, $i=1,\dots r$ of the Cartan algebra. It is natural therefore to call the $r$ monopoles whose magnetic charges equal simple coroots, $g=\av_i$, simple monopoles. Similarly the electric charges, which equal the simple coroots $q=\av_i$, will be referred to as simple electric charges.

\begin{figure}[tbh]
  \centering \includegraphics[ height=3cm, keepaspectratio=true, angle=0]{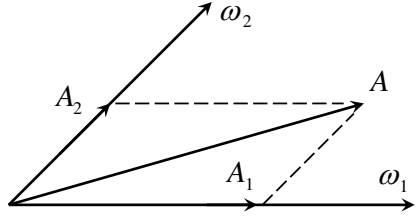}
\vspace{0cm}
\caption{ The vector $A$ representing the VEV of the scalar field in the basis of fundamental weights $\omega_1,\,\omega_2$  for SO(5) gauge group, compare Fig. \ref{SO5}.
 }\label{SO5-A} 
   \end{figure}

Witten and Olive \cite{Witten:1978mh} argued that for each massive dyon 
available in the theory there exists a related central charge $\CZ_{(q,g)}$
\begin{equation}
 \CZ_{(g,q)}     \,=\,   g \cdot \AD+q\cdot A ~,
\label{Z}
	\end{equation}
where $g$  and $q$ are the magnetic and electric charges of the dyon, and that this central charge defines the dyon mass
$ m_{(g,q)}= 2^{1/2} |\,\CZ_{(g,q)}\,|$. To make Eq. (\ref{Z}) more transparent it is convenient following \cite{Kuchiev:2008mv} to present the scalar and dual fields in the basis of fundamental weights
$\omega_i$
\begin{equation}
A\,~=\,~\sum_{i=1}^r \,A_i~\omega_i~,
\quad\quad\quad\quad
 A_\text{\it D}\,=\,\sum_{i=1}^r\,A_{D,\,i}~\omega_i~,
\label{ADi}
\end{equation}
where $A_i$ and $A_{D,\,i}$ are the expansion coefficients, see
Fig. \ref{SO5-A}.
   \noindent
The usefulness of this basis follows from the orthogonal condition between the simple coroots and fundamental weights of the Cartan algebra
\begin{equation}
\av_i\cdot\omega_j\,=\,\delta_{ij}~,
\label{avo}
\end{equation}
which reduces Eq. (\ref{Z}) to a transparent form
\begin{equation}
\CZ_{\,\CG}   \,=\,  \sum_{i=1}^r\,\big(\, n^{(g)}_i \, A_{D,\,i} +n^{(q)}_i\, A_i\big) ~.
\label{Znn}
\end{equation}
Here integers $n^{(g)}_i,\,n^{(q)}_i$ are from Eqs. (\ref{g=ng}). 
Eq. (\ref{Znn}) shows that 
$A_i$ and $A_{D,\,i}$ have a clear physical meaning.
They equal the central charges for simple electric charges and  simple monopoles respectively, $\CZ_{\,(0,\av_i)}=A_i$, $\CZ_{(\av_i,0)}=A_{D,\,i}$.

\subsection{Permutations of simple charges}

 Eq. (\ref{Znn}) makes it convenient to use the sets of the coefficients $A_i$ and $A_{D,\,i}$ (aka the central charges for simple electric charges and  simple monopoles) as arguments of the superpotential $\CF(A)=\CF(A_1,\dots A_r)$ and its dual $\CF_D(A_D)=\CF_D(A_{D,1},\dots A_{D,\,r})$. 
In the following discussion it is presumed that $A$ belongs to the main Weyl chamber, which means that $A_i > 0$.

\begin{figure}[tbh]
  \centering \includegraphics[ height=3.8cm, keepaspectratio=true, angle=0]{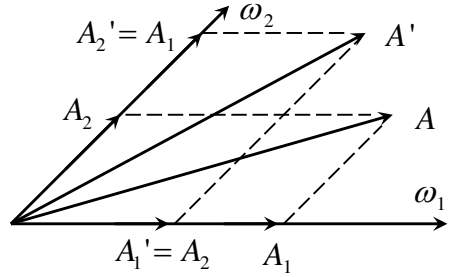}
\vspace{0cm}
\caption{The permutation $(A_1,A_2)\rightarrow 
(A_1^{\,\prime},A_2^{\,\prime})=
(A_2,A_1)$ for the theory with SO(5) gauge group; Eq.(\ref{Fsym}) gives $\CF(A)=\CF(A^{\,\prime})$; notation is same as in Fig. \ref{SO5-A}.
}
\label{SO5-Sr-A} 
   \end{figure}

Let us verify that the function  $\CF(A)=\CF(A_1,\dots A_r)$ is symmetric under an arbitrary  permutation of its arguments
\begin{align}
&\CF(A_1^{\prime},\dots A_r^{\prime})=\CF(A_1,\dots A_r),
\label{Fsym}
\\
&A_i^{\,\prime}\,=\,\sum_{j=1}^r{\mathcal P}_{ij}\,A_j ~,
\label{perm}
\end{align}
Here ${\mathcal P}$ is the $r\times r$ matrix that describes a permutation of $r$ objects. This matrix has only one nonzero matrix element in each column and each line, which equals unity, while the matrix is not degenerate; these conditions imply that it is orthogonal, ${\mathcal P}^T{\mathcal P}=1$. Eqs. (\ref{Fsym}), (\ref{perm}) are illustrated in Fig. \ref{SO5-Sr-A} for SO(5) gauge group. 
   \noindent

To prove the symmetry condition (\ref{Fsym}) one needs to make three steps. First, notice that the permutation of the components of the scalar field in (\ref{perm})
results in a similar permutation of the components of the dual field
\begin{equation}
A_{D,\,i}^{\,\prime}\,=\,\sum_{j=1}^r{\mathcal P}_{ij}\,A_{D,\,j} ~.
\label{permD}
\end{equation}
For strong fields the later relation is justified by the perturbation theory, which guarantees that
\begin{equation}
 A_\text{\it D}\,\approx\,(2\pi)^{-1}\,i\,\hv\, A\,\ln (\,A^2/\Lambda^2)~,
\label{ADPT}
\end{equation}
where  $\hv$ is the dual Coxeter number of the gauge algebra, which is related to the eigenvalue of the quadratic Casimir operator $C_2$ in the adjoint representation, $2\hv=C_2$,
while the logarithmic factor is large, $\ln |A^2/ \Lambda^2|\gg 1$. Presuming also that $A$ is not close to a wall of the Weyl chamber, $|A_i^2|\sim |A^2|$, one finds
$\ln A^2/ \Lambda^2 \approx \ln A^{\,\prime\,^2}/ \Lambda^2$. 
Hence Eq. (\ref{ADPT}) ascertains that Eq. (\ref{perm}) is necessarily  accompanied by (\ref{permD}). Since the theory is governed by holomorphic functions while the considered symmetry group is discrete, one can extend the validity of this statement to arbitrary coupling claiming that the permutation of the set $A_i$ in Eq.(\ref{perm}) necessitates the similar permutation (\ref{permD}) for the set $A_{D,\,i}$.

Second, observe that under the transformation of the fields described in Eqs. (\ref{perm}) and (\ref{permD}) the full set of central charges of all dyons does not change.  Since central charges identify the $\CN=2$ supersymmetric transformations one can be certain that the invariance of the set of central charges implies that the system is symmetric under the transformation of the fields  (\ref{perm}), (\ref{permD}).

Third and last, the symmetry of the system under (\ref{perm}), (\ref{permD}) implies that the superpotential belongs to some representation of the corresponding group of permutations $S_r$. 
Clearly it should be an irreducible representation.  Hence the symmetry condition (\ref{Fsym}) is the only option because any other irreducible representation would make the theory inadequate at weak couplings. 

This discussion justifies the validity of Eq.(\ref{Fsym}).
Using similar arguments one finds that the dual superpotential is also a symmetric function
\begin{equation}
\CF(A_{D,\,1}^{\prime},\dots A_{D,\,r}^{\prime})=\CF(A_{D,\,1},\dots A_{D,\,r}),
\label{FDsym}
\end{equation}
where $A_{D,\,i}^{\prime}$ are from (\ref{permD}). For more detailed discussion see   \cite{Kuchiev:2011vh}. 

We conclude that the theory possesses the symmetry under the group of permutations $S_r$, which manifests itself via Eqs. (\ref{Fsym}) and (\ref{FDsym}).
Remember that $A_i$ and $A_{D,\,i}$ represent the central charges of simple electric charges and simple monopoles. Eqs.(\ref{perm}), (\ref{permD}) can therefore be described as a simultaneous permutation of simple electric charges, which is accompanied by the simultaneous permutation of simple magnetic charges.

\section{Manifestations of symmetry under $S_r$}
\subsection{Superpotential, scalar field and $\tau$ matrix} 
For the SU(2) gauge symmetry the duality of the supersymmetric $\CN=2$ theory introduced in
\cite{Seiberg:1994rs}  can be described 
via a Legendre-type transformation $\CF(A)\rightarrow \CF_D(A_D)=\CF(A)-AA_D$, in which the product $\Xi= AA_D$ plays the role of a generating function for this transform. 
To extend the dual conditions to the theory with an arbitrary gauge symmetry one
can follow a similar approach, considering the Legendre-type transform in which
the generating function $\Xi$ is a bilinear form of the fields $A$ and $A_D$. 
The symmetry conditions (\ref{Fsym}) and (\ref{FDsym})   allow only one option,  for which the bilinear is nondegenerate, $\Xi=\sum_{i=1}^r A_iA_{D,\,i}$.
Hence we find
\begin{equation}
\CF_D(A_D)\,=\, \CF(A)-\sum_{i=1}^{r} A_iA_{D,\,i}~. 
\label{Fdual}
\end{equation}
Differentiating this identity  we derive
\begin{equation}
A_{D,\,i}\,=\frac{\partial \CF}{\partial A_i}~,
\quad\quad\quad
A_{i}~=-\frac{\partial \CF_D}{\partial A_{D,\,i}}~.
\label{AD=dF/dA}
\end{equation}
Observe that the basis of the fundamental weight, which is used for $A$ and $A_D$,
makes these relations very transparent. 
When they are presented in any other conventional basis, say basis of simple roots or coroots, or the orthogonal basis {\it etc}, these relations would include additional geometric factors, which specify the chosen basis. Another notable point is that Refs. (\ref{AD=dF/dA}) are formulated entirely in terms of central charges for simple electric charges and simple monopoles (remember $A_i$ and $A_{D,\,i}$ are these central charges). 
Thus these central charges represent the full set of variables, which are needed to formulate the theory.

Consider now the $\tau$-matrix of the coupling constants. In the basis of the fundamental weights its matrix elements are
\begin{equation}
\tau_{ij}(A)\,=\,
\frac{\partial A_{D,\,i} }{\partial A_{j} }\,=\,
\frac{\partial^2 \CF(A)}{\partial A_{i}\,\partial A_{j}}~.
\label{tau}
\end{equation}
The invariance  of the superpotential 
under $S_r$ in (\ref{Fsym}) implies the following transformation of $\tau(A)$ 
\begin{equation}
\tau(A^{\,\prime})\,=\,{\cal P}\,\tau(A)\,{\cal P}^{\,T}~,
\label{Ptau}
\end{equation}
where $\mathcal P$ is the matrix, which defines a transformation $A \rightarrow A^{\,\prime}$ in (\ref{perm}) (to avoid confusion let us repeat, both $\mathcal P$ and $\tau(A)$ are taken in the basis of the fundamental weights).

\subsection{Large scalar field}
At large scalar field, $|A^2|\gg |\Lambda^2|$, the coupling is weak and we find from (\ref{ADPT})
\begin{equation}
\tau_{ij}\,\approx\,\delta_{ij}\,\frac{i}{2\pi}\,\hv \,\ln\frac{A^2}{\Lambda^2}
\,\approx\,\delta_{ij}\,\frac{i}{2\pi}\,\hv \,\ln\frac{\sum_k A_k^2}{\Lambda^2}~.
\label{tauPTh}
\end{equation}
Here in the last identity the logarithmic function is rewritten to make it explicitly invariant under $S_r$, $\ln A^2\approx \ln \big(\sum_kA_k^2\big)$,
which is possible since the logarithm is presumed large while $A$ not close to walls of the Weyl chamber. 
Observe that at weak coupling the $\tau$-matrix is diagonal only in the basis of the fundamental weights. For any other conventional choice of the basis (simple roots, orthogonal basis {\it etc}) it is not diagonal. 

Recovering the superpotential at weak coupling from Eqs. (\ref{tau}) and (\ref{tauPTh}) one finds 
\begin{equation}
\CF(A)\,\approx\,\frac{i}{4\pi}\,\hv\,\sum_{i=1}^r \,A_i^2\,\ln\frac{\sum_j A_j^2}{\Lambda^2}~.
\label{Fweak}
\end{equation}
Clearly it is explicitly invariant under permutations $\mathcal{P}$ from (\ref{perm}). Observe once again that the basis of fundamental weights makes essential properties of the theory  transparent.

Compare Eq.(\ref{Fweak}) with the known expression for the superpotential at weak coupling
\begin{equation}
\CF_0(A)\,\approx\,\frac{i}{8\pi}\sum_\alpha \,(A\cdot \alpha)^2\ln\frac{A^2}{\Lambda^2}
\,\approx\, \frac{i}{4\pi} \,\hv\,A^2\ln\frac{A^2}{\Lambda^2}~.
\label{Ftrad}
\end{equation}
The summation in the middle expression here runs over all roots.
To derive the final result one takes into account the identity
$\sum_\alpha\alpha \otimes \alpha =2\hv$,
and presumes that since the logarithm is large and $A$ is not close to a wall of the Weyl chamber the logarithmic factor is a smooth function of $A$, and as such can be taken out of summation.
Note  an important distinction separating $\CF(A)$ from $\CF_0(A)$ in Eqs. (\ref{Fweak}) and (\ref{Ftrad}), respectively. 
The superpotential $\CF(A)$ incorporates the factor $\sum_{i=1}^r A_i^2$, which is symmetric under the permutations from $S_r$ defined in (\ref{Fsym}). 
In contrast $\CF_0(A)$ possesses instead a factor $A^2$,
which does not comply with symmetry under $S_r$. 
In line with the arguments of the present work, Eq. (\ref{Ftrad}) is 
inadequate for the low-energy problem. 

Qualitatively this situation can be compared with the elastic scattering of X-rays in crystals. The interaction between a photon and each Wigner-Seitz cell (small part of the crystal) is weak. However, the angular distribution of X-rays is governed by the symmetry of the crystal reciprocal lattice, which therefore represents the symmetry of this problem.  
An important for us lesson is that the presence of the lattice can change the 
symmetry properties of the system even when the naive perturbation theory, which 
describes the interaction with each small part of the lattice, is applicable.

This work argues that a similar phenomenon takes place in the supersymmetric $\CN=2$ gauge theory. The lattice $\mathbb Q^\vee$ of electric and magnetic charges, which is present here, imposes on the system the particular symmetry, which manifests itself in the superpotential $\CF(A)$ in Eq.(\ref{Fweak}). 
It is important that the phenomena described by this superpotential are related to the low-energy region, $\varepsilon^2 \ll |A^2|$, where the electric charges and monopoles  are well defined.
At higher energies (smaller distances) the description with the help of these charges is inadequate. In contrast, the superpotential $\CF_0(A)$ in (\ref{Ftrad}) follows from the conventional perturbation theory based on the one-loop calculations. 
It reliably describes events at high energy,
$\varepsilon^2 > |A^2|$, but may not be sufficiently equipped for reproducing the effects related to electric charges and monopoles, which manifest themselves at lower energies. 
One can anticipate therefore that the conventional perturbative approach may not reproduce the symmetry conditions imposed by the lattice $\mathbb Q^\vee$ of these charges. 

These simple, qualitative arguments show that the discrepancy between $\CF_0(A)$ in (\ref{Ftrad}) and $\CF(A)$ in Eq.(\ref{Fweak}) is not  surprising, and that 
for low energies the latter is more trustworthy.

\subsection{Weyl vector alignment}
\label{Degeneracy-Weyl}

\begin{figure}[tbh]
  \centering \includegraphics[ height=6.5cm, keepaspectratio=true, angle=0]{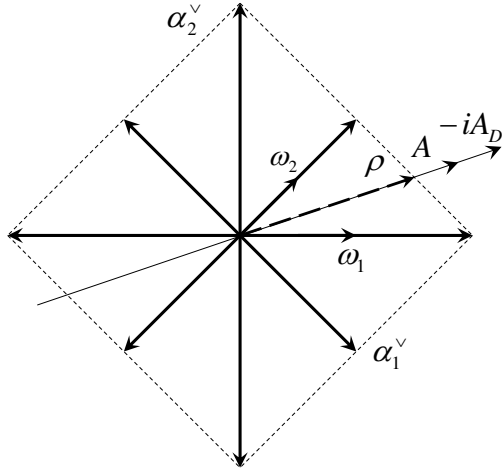}
\vspace{0cm}
\caption{The Weyl vector alignment: if one of the vectors $A$ or $A_D$, which represent the scalar and dual fields, is aligned along the Weyl vector $\rho$, then the other is aligned along the same vector as well (\ref{Weyl}); notation is same as in Figs. \ref{SO5} and \ref{SO5-A},  for simplicity of presentation it is presumed that both $A$ and $-iA_D$ are real.
 }\label{SO5-AAD} 
   \end{figure}

Consider the special case when the scalar field is chosen to be parallel to the Weyl vector $\rho$ of the Cartan algebra. Remember that this vector is defined as follows
\begin{equation}
\rho\,=\,\sum_{i=1}^r \,\omega_i~,
\label{ro}
\end{equation}
see Fig. \ref{SO5} for illustration. 
Remarkably, the dual field $A_D$ in this case proves to be aligned along same direction.  The reverse is also valid, i. e. the alignment of $A_D$ along the Weyl vector implies the alignment of $A$ along this direction.
This property, which can be called the Weyl vector alignment, can be presented as follows
\begin{equation}
A\,=k\,\rho \quad \Longleftrightarrow \quad A_D \,=\,k_D\,\rho~,
\label{Weyl}
\end{equation}
where $k$ and $k_D$ are numbers. 
The double arrow here indicates that one identity implies the other.
Fig. \ref{SO5-AAD} illustrates this configuration of the fields for SO(5) gauge group. 

To verify validity of (\ref{Weyl}) note that $A=k\,\rho$ implies that all $A_i$ are same, $A_i=k$, $i=1,\dots r$. Therefore Eq.(\ref{AD=dF/dA}) gives $A_{D,\,i}=\partial \CF(k,\,\dots k)/\partial A_i$. Since we know that  the superpotential is symmetric (\ref{FDsym}), we find that all $A_{D,\,i}$ are also same, $A_{D,\,i}=\partial \CF(k,\,\dots k)/\partial A_1\equiv k_D$. 
The last identity implies $A_D=k_D \,\rho$, which proves Eq.(\ref{Weyl}) when it is read from left to right; its validity in the opposite direction is verified similarly.
   \noindent

As mentioned, the alignment $A=k \,\rho$ means that $A_i=k$ and we find from (\ref{Znn}) that the central charges of all 
simple electric charges 
are degenerate $\CZ_{(0,\av_i)}=k$.  Similarly, $A_D=k_D \,\rho$ implies that
the central charges of simple monopoles are also degenerate
$\CZ_{(\av_i,0)}=A_{D,\,i}=k_D$.
Thus in physical terms the Weyl vector alignment (\ref{Weyl}) means 
that the degeneracy of central charges for simple electric charges implies the degeneracy of central charges for simple monopoles, and vice versa.

This statement can be rephrased for masses.
Note that $A_i=k$ implies that the masses of simple electric charges are degenerate, $m_{(0,\av_i)}= 2^{1/2} |\CZ_{(0,av_i)}|=2^{1/2}k$.
Remember that we assume that $A$ is in the main Weyl chamber, which means that $A_i\ge 0$. Consequently, we can start the argument from the masses saying that
if the masses of simple electric charges are degenerate, $m_{(0,\av_i)}=2^{1/2}k$, then their central charges are also same $A_i=k$. The previous discussion ensures that in this case the central charges of simple monopoles are
also same, $A_{D,\,i}=k_D$.  As a result we find that the masses of simple monopoles are degenerate $m_{(\av_i)}=2^{1/2}|k_D|$. 
Summarizing we find that
\begin{equation}
m_{(0,\av_i)}\,=\sqrt 2 \,|k|
\quad \quad
\Longleftrightarrow 
\quad \quad
m_{(\av_i,0)}\,=\,\sqrt 2 \,|k_D|~.
\label{Weylm}
\end{equation}
The arguments above support this statement  when it is read following the arrow from left to right; its validity in the opposite direction, from right to left, is verified similarly.
Eq. (\ref{Weylm}) shows that the degeneracy of masses of simple electric charges implies the degeneracy of masses of simple monopoles, and vice versa,
which fits well within the general idea of duality.
This fact supports the validity of (\ref{Weylm}).

\subsection{Massless monopoles}
\begin{figure}[tbh]
  \centering \includegraphics[ height=2.5cm, keepaspectratio=true, angle=1]{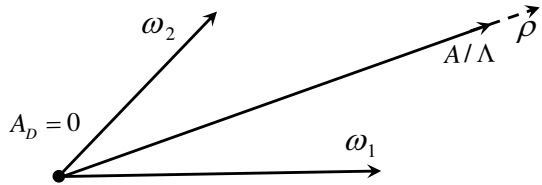}
\vspace{0cm}
\caption{The corollary to the Weyl vector alignment:  if $A_D=0$ and all monopoles are massless then the vector $A$ is aligned with the Weyl vector $\rho$ (\ref{corollary}); notation is the same as in Figs. \ref{SO5-A} and \ref{SO5-AAD}, but $A$ is scaled by the cut off parameter $\Lambda$.
 }\label{SO5-AD-zero} 
   \end{figure}

An interesting implication of the Weyl vector alignment 
arises in the limit $A_\text{\it D}= 0$, when all monopoles are massless.
Eq. (\ref{Weyl}) shows that in this case 
\begin{equation}
A_\text{\it D}=0\quad\Longrightarrow \quad A\,=\, c \,\Lambda\, \rho~.
\label{corollary}
\end{equation}
Here $\Lambda$ is a cutoff parameter of the theory, which is written
on the basis of simple dimensional counting, while $c$ is a number, which
can be  evaluated using the model of \cite{Kuchiev:2009ez}. 
Fig. \ref{SO5-AD-zero} illustrates Eq.(\ref{corollary})  for SO(5) gauge theory.
   \noindent

Using (\ref{Weylm})  one can rephrase (\ref{corollary}) by saying that when the masses of simple monopoles are zero, the masses of simple electric charges 
are degenerate $m_{(0,\av_i)}=c\,2^{1/2}\,|\Lambda|$.
The point where the dual field turns zero plays an exceptionally important role in the theory. In particular, it was pointed out in \cite{Seiberg:1994rs} that here the $\CN=2$ supersymmetric gauge theory can be broken down explicitly to the case of $\CN=1$ supersymmetry. It is interesting therefore that (\ref{corollary}), which
specifies the field $A$ related to this point, follows directly
from the symmetry condition (\ref{Fsym}) and does not rely 
on any additional calculations or model approximations. 

\section {Simple nature of general solution}
\label{simple}
The discussion given above implies that the solution of the Sieberg-Witten problem for the supersymmetric $\CN=2$ gauge theory with a general gauge group turns out to be much simpler than one could have anticipated. There are only two parameters, the rank $r$ and the dual Coxeter number $\hv$ of the gauge group, which govern dynamics of the system. The first, obviously, defines the number of arguments of the superpotential, while the second governs the weak coupling behavior in (\ref{ADPT}). All other parameters related to the 
detailed structure of the Cartan matrix of the gauge group prove irrelevant. It is immaterial,
in particular,  whether the gauge group is simply laced or not, belongs to one classical series or another, 
or  is an exceptional one. 

This claim was previously articulated in \cite{Kuchiev:2011vh} on the basis of the model proposed there. The present discussion can be used to support this statement from a different perspective. The symmetry of the superpotential in (\ref{Fsym}) (and in (\ref{FDsym})) simply leaves no room for any particulars related to the Cartan matrix. 
Any condition on any given argument $A_i$ in $\CF(A)$ (or $A_{D,\,i}$ in $\CF_D(A_D)$), which arises from the discrete symmetries, boundary conditions, and monodromies of the superpotential is equally applied to any other argument of this symmetric function, which ensures that the superpotential is independent on the detailed structure of Cartan's matrix.

\section{Summary}
\label{Summary}

It is shown that the superpotential, which provides the low-energy description of the supersymmetric $\CN=2$  gauge theory, obeys a symmetry condition (\ref{Fsym}). In simple physical terms it can be described as a symmetry under an arbitrary permutation of $r$ simple electric charges, which is accompanied by the simultaneous identical permutation of $r$ simple monopoles. This symmetry condition implies that the theory can be conveniently presented in terms of central charges of simple monopoles and simple electric charges,
see (\ref{AD=dF/dA}). 

Another interesting implication is the Weyl vector alignment (\ref{Weyl}), which ensures that the degeneracy in the spectrum of the monopole masses is closely related to degeneracy of masses of electric charges. 
Its corollary (\ref{corollary}) defines the location of the important state of the theory where all monopoles are massless. 
These properties have an elegant geometrical representation when expressed via the Weyl vector of the Cartan algebra.


The found discrete symmetry simplifies the theory and  can be used to support the validity of the solution of the Seiberg-Witten problem for a theory with an arbitrary gauge group proposed previously in \cite{Kuchiev:2009ez}, see discussion in \cite{Kuchiev:2011vh}.

%

\vspace{0.3cm}
The financial support of the Australian Research Council is acknowledged.

\end{document}